\documentclass{osa-article}

\journal{boe}

\articletype{Research Article}

\usepackage{rebuttal}
\usepackage{textgreek}
\usepackage{float}
\newfloat{algorithm}{h}{algfl}
\floatname{algorithm}{Algorithm}

\DeclareGraphicsExtensions{.ps}

\renewcommand{\vec}[1]{\mathbf{#1}}

\newcommand{\pars}[1]{\left({#1}\right)}
\newcommand{\R}{\mathbb{R}}

\newcommand{\period}{\;\text{.}}
\newcommand{\comma}{\;\text{,}}
\newcommand{\alg}{\text{alg}}
\DeclareMathOperator*{\argmin}{arg\,min}
\DeclareMathOperator*{\argmax}{arg\,max}

\DeclareMathOperator{\Prob}{P}

\DeclareMathOperator{\Li}{L}
\DeclareMathOperator{\X}{\vec{X}}
\DeclareMathOperator{\Y}{\vec{Y}}
\DeclareMathOperator{\Reg}{R}
\DeclareMathOperator{\AD}{AD}

\DeclareMathOperator{\IFV}{IFV}

\begin{document}

\title{Maximum A Posteriori Signal Recovery for Optical Coherence Tomography Angiography Image Generation and Denoising}

\author{Lennart Husvogt,\authormark{1,2*} Stefan B Ploner,\authormark{1} Siyu Chen,\authormark{2} Daniel~Stromer,\authormark{1,2} Julia Schottenhamml,\authormark{1} 
A~Yasin~Alibhai,\authormark{3} Eric~Moult,\authormark{2} Nadia~K~Waheed,\authormark{3} James~G~Fujimoto,\authormark{2} and Andreas Maier\authormark{1}}

\address{\authormark{1}Pattern Recognition Lab, Friedrich-Alexander-Universit\"at Erlangen-N\"urnberg, Germany\\
\authormark{2}Research Laboratory of Electronics, Massachusetts Institute of Technology, Cambridge, USA\\
\authormark{3}New England Eye Center, Tufts School of Medicine, Boston, USA}

\email{\authormark{*}lennart.husvogt@fau.de} 



\begin{abstract}
Optical coherence tomography angiography (OCTA) is a novel and clinically promising imaging modality to image retinal and sub-retinal vasculature. Based on repeated optical coherence tomography (OCT) scans, intensity changes are observed over time and used to compute OCTA image data. OCTA data are prone to noise and artifacts caused by variations in flow speed and patient movement. We propose a novel iterative maximum a posteriori signal recovery algorithm in order to generate OCTA volumes with reduced noise and increased image quality. This algorithm is based on previous work on probabilistic OCTA signal models and maximum likelihood estimates. Reconstruction results using total variation minimization and wavelet shrinkage for regularization were compared against an OCTA ground truth volume, merged from six co-registered single OCTA volumes. The results show a significant improvement in peak signal-to-noise ratio and structural similarity. The presented algorithm brings together OCTA image generation and \replaced{M2.1}{\link{C2}}{compressed sensing}{Bayesian statistics} and can be developed into new OCTA image generation and denoising algorithms.
\end{abstract}

\section{Introduction}
Over the last years, optical coherence tomography angiography (OCTA) has become increasingly popular in research and clinical imaging. It allows micrometer-resolution imaging of retinal vessels and the choriocapillaris, the latter which is challenging to image using other existing methods \cite{Choi2015}. In addition to that, OCTA is non-invasive compared to other methods, such as fluorescein angiography, which requires the injection of a contrast agent and carries the risk of anaphylactic shock \cite{Novais2016}. OCTA is in clinical use and available in commercial devices by various manufacturers.

Two commonly used methods to generate OCTA data are amplitude decorrelation (AD) and speckle variance (SV). In both cases, optical coherence tomography (OCT) B-scans of the same location are acquired in immediate succession and changes in intensity over time serve as the basis on which OCTA is computed \cite{Jia2012, Mariampillai2008}. \replaced{M1.1}{\link{C1}}{To the best of our knowledge, denoising of OCTA images is usually performed by applying basic filters such as a median filter and background thresholding. Thus, there is considerable opportunity to find new ways of reconstructing OCTA data to improve image quality.}{To the best of our knowledge, denoising of OCTA images is usually performed by applying basic filters such as a median filter and background thresholding \emph{to the measured data}. While these approaches already achieve significant improvement with simple methodology, their underlying assumptions do not accurately apply to the OCTA image formation process. This discrepancy leaves plenty of room for further improvement of OCTA image quality without imposing changes to the acquisition protocol or the device.}

\replaced{M1.2}{\link{C1}}{Reconstruction of medical image data has made substantial progress in recent years and there is a multitude of reconstruction algorithms in computed tomography (CT) and magnetic resonance imaging (MRI). A data consistency term is minimized while regularizers enforce certain properties in the reconstructed data. This, in the case of inverse linear problems, is often coupled with compressed sensing (CS). CS has led to several advances in medical imaging over the last years [5]. These approaches allowed significant reductions in imaging time for MRI and reduction in applied X-ray doses in CT [6-11]. Because OCTA computations are non-linear and non-invertible, we suggest to use Bayesian statistics to develop a data consistency term for the reconstruction of OCTA. This approach can be related to CS in conjunction with a regularizer that enforces sparsity.

Bayesian statistics, such as maximum a posteriori (MAP) estimates, have been used before to reduce speckle noise in OCT images, while CS has already been used in the past for the generation of structural OCT images [12]. Liu et al.\@ and Mohan et al.\@ demonstrated the use of CS for spectral domain OCT to reduce the amount of k-space samples that need to be acquired [13-15]. Young et al.\@ employed CS for real-time volumetric imaging [16] while Fang et al.\@ performed so called multiscale sparsity-based tomographic denoising [17, 18]. This CS method allows the denoising of spectral domain OCT data using custom scan patterns. Zhang et al.\@ also utilized CS by using a linear mask in k-space for subsampling which reduces the number of samples necessary for OCT signal generation [19]. Xu et al.\@ implemented a denosing method for spectral domain OCT also based on CS and achieved real-time capability [20-22].

In this paper we present, to the best of our knowledge, the \emph{first MAP image reconstruction algorithm for OCTA}.
}{
Reconstruction of medical image data has made substantial progress in recent years, which was achieved by inverse problem formulations, which map the measurements backwards through the image formation process to the true observed specimen. Such problems can be solved via iterative reconstruction, where the result image is found by optimizing an objective function comprised of two major parts: First, a data consistency term that ensures that the reconstructed data closely represents the inversely mapped measurements. Secondly, regularization terms are used to improve image quality by enforcing certain properties. Here, the formulation as inverse problem has the crucial advantage that such constraints can be enforced not only on the measurements, but also directly in the reconstructed data. Applications of this principle range through the entire field of medical imaging: In magnetic resonance imaging (MRI), compressed sensing-based regularization is used to enforce sparseness of the reconstructed data, achieving great reductions in imaging time \cite{Lustig2007}. Closely related, in computed tomography (CT), homogeneity within tissue classes was modeled by regularizers, achieving significant dose reduction at equal image quality \cite{Sidky2008, Wu2012, Amrehn2015, Stromer2016, Huang2016}.

A common way for setting up the objective function is to use Bayesian statistics, which has been used in the context of OCT before to reduce speckle noise \cite{Wong2010}. More recently, Ploner et al.\@ introduced a Bayesian signal model for OCTA which uses maximum likelihood estimation (MLE) \cite{Ploner2018}. While this initial model provided a theoretical basis for the data consistency term, image quality was not improved, because the model still lacks the key advantage of the inverse problem formulation: the inclusion of prior information on the reconstructed signal via regularization terms. This would extend the MLE to a complete Bayesian maximum a posteriori (MAP) estimation. In this paper we present, to the best of our knowledge, the first MAP image reconstruction algorithm for OCTA: we extend the OCTA MLE model by Ploner et al.\@ to a MAP estimate and use wavelet shrinkage and total variation minimization as regularizers.}

The evaluation of an algorithm like this poses some challenges though. OCT scans suffer from a wide range of motion artifacts because scan acquisition can take several seconds and patient eyes exhibit involuntary eye motion. This causes severe difficulties in acquiring any ground truth (GT) data that exhibit few motion artifacts and which have sufficient overlap with test data. To address this problem, we generated a low-noise GT OCTA volume by co-registering and merging six individual OCTA volumes with each OCTA volume generated from scans using ten B-scan repetitions each. One of the six volumes also serves as test volume. Because of the co-registration with the GT, reconstruction results can be deformed to match it and error metrics between the two can be computed.

This paper is structured as follows: in the next section we provide a description of the MAP estimate\replaced{M1.3}{\link{C1}}{and our algorithm}{, comprised of the MLE it is based on and the newly introduced regularizers, and the reconstruction algorithm}. It is followed by the experiment section that explains how the OCTA ground truth was generated. Results and discussion sections follow and the paper ends with a conclusion.

\section{Method}
\replaced{M1.4}{\link{C1}}{The reconstruction is based on a MAP estimate, which is described first. The OCTA signal model used within the estimate is described afterwards, followed by the reconstruction algorithm.}{The reconstruction is based on a MAP estimation, which is introduced first. The OCTA MLE signal model and regularizers are described afterwards, followed by the reconstruction algorithm.}

\changed{M2.2}{\link{C2}}{\subsection{Maximum A Posteriori Probability Estimation}}\label{sec:map}
\replaced{M2.3}{\link{C2}}{CS}{Image reconstruction} arose from the need to reconstruct some data $\vec{X}$ from a limited number of measurements $\vec{Y}$. \replaced{M2.4}{\link{C2}}{Specifically}{For instance}, where a linear sampling operation $\vec{A}\in\R^{M\times N}$ maps from a finite signal $\vec{X}\in\R^N$ to
\begin{align}
    \vec{Y} = \vec{A}\vec{X} + \vec{e}\comma{}\label{eq:cs}
\end{align}
with $\vec{Y}\in\R^M$, while $\vec{e}\in\R^M$ models the impact noise has on the measured signal $\vec{Y}$. $\vec{A}$ models the sampling of $\vec{X}$ to the observed signal $\vec{Y}$ and is often called the \emph{system matrix}. The goal is to reconstruct $\vec{X}$ based on the measurements $\vec{Y}$ which is an \emph{inverse problem}. \deleted{M2.5}{\link{C2}}{The challenge which CS addresses, is the case where $\vec{A}$ is a linear mapping with $M<N$. In that case, $\vec{A}$ is being underdetermined which leads to an infinite number of possible solutions for $\vec{X}$. CS allows to find solutions in this case by exploiting sparsity of $\vec{X}$ in a certain basis. A good example for a different basis used in CS are the coefficients of a wavelet transform. $\vec{X}$ is then called $k$-sparse when it has $k$ non-zero entries in that basis.} One approach to solve this inverse problem in the case of OCTA, is to treat it as a MAP estimate \cite{Zeng2010}. By interpreting $\vec{X}$ and $\vec{Y}$ as random variables with their own distributions, the MAP estimate has the goal to find an OCTA volume $\mathbf{\hat{X}}$ which maximizes its probability given the OCT scan $\Y$. This is defined as
\begin{align}
    \mathbf{\hat{X}} =
    \argmax_{\X} \Prob\left(\X \middle| \Y\right) = \argmax_{\X} \frac{\Prob\left(\Y \middle| \X\right)\Prob\pars{\X}}{\Prob\pars{\Y}} = \argmax_{\X} \Prob\left(\Y \middle| \X\right)\Prob\pars{\X} \period \label{eq:map}
\end{align}
It is possible to ignore $\Prob\pars{\Y}$ because it does not depend on $\X$. $\Prob\left(\Y \middle| \X\right)$ is the \replaced{M3.1}{}{\emph{maximum likelihood estimate}}{MLE} for $\Y$ and $\Prob\pars{\X}$ is the \emph{prior probability} of $\X$. If $\Prob\left(\Y \middle| \X\right)$ is a multivariate Gaussian distribution with mean $\mu = \vec{A}\vec{X}$ and covariance matrix $\Sigma = \vec{I}$, maximizing the log probability yields
\begin{align}
    \argmax_{\X} \ln \Prob\left(\Y \middle| \X\right)\Prob\pars{\X} &= \argmax_{\X} \ln \left[ \exp \pars{-\frac{1}{2} \pars{\Y - \vec{A}\X}^T \Sigma^{-1} \pars{\Y - \vec{A}\X}} \lvert \Sigma \rvert^{-\frac{1}{2}} \Prob\pars{\X} \right]\\ 
    &= \argmax_{\X} -\frac{1}{2} \pars{\Y - \vec{A}\X}^T \pars{\Y - \vec{A}\X} + \ln \Prob\pars{\X} \\
    &= \argmin_{\X} \frac{1}{2} \lVert \Y - \vec{A}\X \rVert_2^2 + \lambda \Reg\pars{\X} \period \label{eq:obj_function}
\end{align}
The MAP \replaced{M2.6}{\link{C2}}{interpretation of CS}{ estimate} yields a least squares problem with a regularizer $\Reg\pars{\X}$ in lieu for $\ln \Prob\pars{\X}$ to include prior knowledge about $\X$ or \replaced{M1.5}{\link{C1}}{to enforce sparsity for CS}{as a regularizer}.
In cases such as described by Manhart et al.\@ and Neukirchen et al.\@ where the system matrix $\vec{A}$ is ill-conditioned and the data dimensions are large, the Landweber iteration was successfully used to find a solution for $\vec{X}$ \cite{Manhart2013, Neukirchen2010}. \deleted{M2.7}{\link{C2}}{In addition to that, based on the iterative hard thresholding algorithm, Blumensath generalized CS for cases in which $\vec{A}$ is non-linear and non-invertible and showed that under certain conditions the Landweber iteration can also be used here [9, 10].}

However, because OCTA computations are non-linear and non-invertible, the above Gaussian model for the maximum log likelihood estimate does not apply. This is where the work of Ploner et al.\@ becomes relevant, in which the authors describe maximum log likelihood estimates for AD, SV, and interframe variance (IFV) which are used as $\Prob\left(\Y \middle| \X \right)$ \cite{Ploner2018}. The next section will summarize the OCTA signal models and \replaced{M3.2}{}{maximum likelihood estimates}{MLE} used for reconstruction.

\subsection{Probabilistic OCTA Signal Models}\label{sec:octa_signal_models}
As noted in the introduction, intensity-based OCTA functions are based on changes in intensity over time for a given voxel. For an OCT scan where each B-scan was repeatedly scanned N times, $y_i$ denotes the measured amplitude for scan repetition $i\in \left[1,2, \ldots, N \right]$. AD is the variance of the pair-wise difference amplitudes $y_i$ and $y_{i+1}$ normalized by
\begin{align}
    \frac{1}{\sqrt{y^2_i + y^2_{i+1}}}\period
\end{align}
We assume that $\mu=0$ for the differences between $y_i$ and $y_{i+1}$ because both are identically distributed. The likelihood to acquire the measurements $y_1$ to $y_N$ is then
\begin{align}
    L_{\AD} = \prod_{i=1}^{N-1} \Prob\left(\frac{y_i - y_{i+1}}{\sqrt{y_i^2+y_{i+1}^2}} \middle| \mu=0, \sigma^2=x\right)
    =\prod_{i=1}^{N-1}\frac{1}{\sqrt{2\pi x}} \exp \left( -\frac{\frac{\left(y_i-y_{i+1}\right)^2}{y_i^2+y_{i+1}^2}}{2x} \right) \period
\end{align}
Using the maximum log-likelihood estimation one obtains
\begin{align}
    \argmax_x \log L_{\AD} = \argmax_x \pars{N-1} \cdot \log \left( \frac{1}{\sqrt{2\pi x}} \right) - \sum_{i=1}^{N-1}\frac{\frac{\pars{y_i-y_{i}}^2}{y_i^2+y^2_{i+1}}}{2x} \comma
\end{align}
with the derivative over $x$ being
\begin{align}
    \frac{d \log L_{\AD}}{d x}=\frac{-(N-1)\cdot x+\sum_{i=1}^{N-1}\frac{\pars{y_i-y_{i+1}}^2}{y_i^2+y_{i+1}^2}}{2x^2} \period \label{eq:deriv_AD}
\end{align}
Setting the derivative to zero and solving for $x$ results in the definition of AD \cite{Jia2012}
\begin{align}
    x = \frac{1}{N-1}\sum_{i=1}^{N-1}\frac{\pars{y_i-y_{i+1}}^2}{y_i^2+y_{i+1}^2} = 1-\frac{1}{N-1}\sum_{i=1}^{N-1}\frac{y_i\cdot y_{i+1}}{\frac{1}{2}\pars{y_i^2+y_{i+1}^2}}\period  \label{eq:ad}
\end{align}

Based on SV, Ploner et al.\@ proposed a new angiography formula, IFV \cite{Ploner2018}. Compared to SV, IFV computes the variance of the differences between voxel values over time but without the normalization used in AD
\begin{align}
    L_{\IFV} = \prod_{i=1}^{N-1} \Prob\left(y_i-y_{i+1} \middle| \mu=0,\sigma^2=x\right)
    =\prod_{i=1}^{N-1}\frac{1}{\sqrt{2\pi x}} \exp\left( -\frac{\pars{y_i-y_{i+1}}^2}{2x} \right)\period
\end{align}
Again, as with AD, the assumption is that $\mu=0$. After applying the log function
\begin{align}
    \log L_{\IFV} = \pars{N-1}\cdot\log \left( \frac{1}{\sqrt{2\pi x}} \right) -\sum_{i=1}^{N-1}\frac{\pars{y_i-y_{i+1}}^2}{2x}\comma
\end{align}
and taking the first degree derivative with respect to $x$
\begin{align}
    \frac{d\log L_{\IFV}}{dx} = - \frac{(N-1)\cdot x+\sum_{i=1}^{N-1}\pars{y_i-y_{i+1}}^2}{2x^2} \comma \label{eq:deriv_IFV}
\end{align}
the maximum can then be found at 
\begin{align}
    x &=\frac{1}{N-1}\sum_{i=1}^{N-1}(y_i-y_{i+1})^2 \period \label{eq:ifv}
\end{align}
The \replaced{M3.3}{}{maximum likelihood estimate}{MLE} for SV can be found in the supplemental document.

\changed{M1.6}{\link{C1}}{By including prior knowledge in form of a regularizer, the MLE can be extended into a MAP estimate. For this paper we evaluated two different regularizers: wavelet shrinkage, and total variation minimization. Wavelet shrinkage works by transforming the OCTA volume into wavelet domain and setting the wavelet coefficients below a certain threshold to zero \cite{Chang2000}. For total variation regularization, in which the variation over the signal is minimized, we used the algorithm by Chambolle et al.\@ \cite{Chambolle2004}.}\changed{M2.8}{\link{C2}}{ These regularizers are also commonly used in compressed sensing to enforce sparsity. The relationship of the algorithm described in this paper to compressed sensing is described in the supplementary.}

\subsection{Reconstruction Algorithm}
\begin{algorithm}[htbp]
    \textbf{Objective:} Find $\mathbf{\hat{X}}=\argmax_{\X} \Prob\left( \X \middle| \Y \right)$\\
    \textbf{Input:} Supply the following components and parameters:
    \begin{itemize}
        \item Step size $\lambda$
        \item Regularizer $\Reg$ and parameters
        \item Total number of iterations $N_{\text{iter}}$ and number of iterations $N_{\text{reg}}$ for the regularizer
    \end{itemize}
    \textbf{Initialization:} Set $\vec{X}_0=\vec{A}_{\alg}\pars{y}$\\
    \textbf{Loop:}
    \begin{itemize}
        \item Update the OCTA volume $\vec{X}_{k+1} = \vec{X}_k + \lambda \nabla\Li_\text{alg}\pars{\vec{X}_k}$.
        \item Every $N_{\text{reg}}$ iterations apply the regularizer $\vec{X}_{k+1} = \Reg\pars{\vec{X}_k}$.
    \end{itemize}
    \textbf{End of Loop:}\\
    \textbf{Result:} The output $\vec{\hat{X}}$
    \caption{Description of the OCTA MAP reconstruction algorithm. The current estimate $ \vec{X}_k$ is iteratively updated using the element-wise derivative $\nabla\Li_\text{alg}\pars{\vec{X}_k}$ while a regularizer is applied every $N_{\text{reg}}$ iterations. The result is $\vec{\hat{X}}$.}
    \label{alg:algorithm_listing}
\end{algorithm}

For the purposes of this paper, both structural and angiography OCT volumes consist of $N_b$ B-scans. A B-scan is a 2D image composed of $N_a$ A-scans, which in turn contain $N_s$ samples or pixels. Furthermore, in order to generate OCTA data, every structural B-scan is scanned $N_r$ times. A structural OCT volume is thus $\vec{Y}\in\R^{N_b\cdot N_r\cdot N_a\cdot N_s}$ while an OCTA volume is $\vec{X}\in\R^{N_b\cdot N_a\cdot N_s}$. 

Even though AD and IFV are non-linear and non-invertible, it is possible to compute an OCTA reconstruction of $\vec{X}$ using the MAP estimate. For this we propose a novel data consistency term, based on the \replaced{M3.4}{}{maximum likelihood estimates}{MLE} in section \ref{sec:octa_signal_models} and the MAP estimate in Eq.~(\ref{eq:map}). The goal is to find an OCTA volume $\vec{\hat{X}}$ that maximizes the probabilities at each voxel position that fit the OCT observations $\vec{Y}$ by maximizing
\begin{align}
    \vec{\hat{X}} &= \argmax_{\vec{X}} \log \left[ \Prob\left(\Y \middle| \X \right) \Prob\pars{\X} \right] \\
    &= \argmax_{\vec{X}} \log\Prob\left(\Y \middle| \X \right) + \Reg \pars{\X} \period
    \label{eq:cs_li_obj_fun}
\end{align}
Using the observations in $\vec{Y}$ we can find the OCTA volume $\vec{\hat{X}}$ that maximizes the posterior probability of $\vec{Y}$. Assuming that $\X$ is uniformly distributed, $\Prob\pars{\X}$ becomes constant, and can be disregarded. A regularizer is added in its place that fulfills a similar function, enforcing prior knowledge about $\vec{X}$. \deleted{M1.7}{\link{C1}}{For this paper we evaluated two different regularizers: wavelet shrinkage, and total variation minimization. Wavelet shrinkage works by transforming the OCTA volume into wavelet domain and setting the wavelet coefficients below a certain threshold to zero [16]. For total variation regularization, in which the variation over the signal is minimized, we used the algorithm by Chambolle et al.\@ [17].} \deleted{M2.9}{\link{C2}}{These regularizers have in common that they introduce sparsity into the OCTA volume, which relates the reconstruction to CS.} Finally, we search $\mathbf{\hat{X}}=\argmax_{\X} \Prob\left( \X \middle| \Y \right)$ using the Landweber iteration 
\begin{align}
    \vec{X}_{k+1} = \vec{X}_k + \lambda \nabla\Li_\text{alg}\pars{\vec{X}_k} \comma
\end{align}
with $\lambda$ being the relaxation factor or step size and $\nabla\Li_\text{alg}$ the element-wise derivative for $\text{alg} \in \{\text{AD}, \text{IFV}\}$ from Eq.~(\ref{eq:deriv_AD}) and Eq.~(\ref{eq:deriv_IFV}) \cite{Landweber1951}. The regularizer is applied to the currently reconstructed volume $\vec{X}_k$ every $N_{reg}$ iterations \cite{Manhart2013, Neukirchen2010}. \figurename~\ref{alg:algorithm_listing} shows the complete reconstruction algorithm, including the required parameters.

\section{Evaluation}
This section describes the experiments, the data used, including the generation of the GT data, and patient data.

\subsection{Experiments}
Two experiments using the GT and patient data were conducted:
\begin{enumerate}
    \item Comparison of reconstruction results with the GT for AD and IFV OCTA formulas using wavelet shrinkage and total variation minimization as regularizers for different numbers of repeated samples.
    \item Qualitative evaluation of two scans of patient's eyes.
\end{enumerate}
The wavelet shrinkage regularizer is used with a threshold of 5e-4 using Haar wavelets for a total of 1000 iterations. The total variation regularizer uses a denoising weight of 1e-4 and ten iterations for the regularizer. The reconstruction was run for 2000 iterations. For the reconstruction of AD data, a step size of $\lambda=5e-6$ was chosen and for IFV data $\lambda=3e-6$. The number of samples in experiment one refers to the number of repeated OCT B-scans used in computing the OCTA signal. The reconstruction and experiments were implemented in Python using NumPy and scikit-learn \cite{VanDerWalt2011, VanDerWalt2014}.

\subsection{Ground Truth Data}
In order to evaluate the performance of our method, we decided to compare the reconstruction results with GT OCTA data. The GT gold standard for structural OCT is averaged B-scans \cite{Mayer2012}. However, patient motion and its associated artifacts pose a significant challenge for the acquisition of a GT. In order to generate GT OCTA data with minimal motion artifacts, we chose the motion correction  algorithm by Kraus and Ploner et al.\@ \cite{Kraus2012, Kraus2014, ploner20183, Ploner2020}. The motion correction algorithm requires at least two OCT volumes which were raster-scanned with orthogonal fast-scan directions. It is based on the principle of non-rigid registration, i.e.\@ the motion correction algorithm estimates a 3D displacement field for each scan such that similarity between the displaced inputs is maximized, while using both OCT and OCTA for registration. After deforming the volumes to match each other, the motion correction algorithm is able to produce a merged and interpolated result by performing a weighted average per A-scan. This allows to correct for gaps and overlaps in the input volumes as long as the missing data is contained in another volume. The more volumes are used as inputs, the higher the signal-to-noise ratio (SNR) of the motion-corrected and merged result, which includes both OCT and OCTA channels \cite{ploner20183, Ploner2020}. It is important to note that the merging step of the motion correction algorithm is only used for generating the GT. Since the chosen test volume is also co-registered with the GT it is possible to compute comparative metrics. The merged GT OCTA volume with high SNR allows to compare the reconstructed results. Although noise is substantively reduced, the GT OCTA volume is not perfect since it still exhibits some low levels of noise. Because of this it is more of a silver than a gold standard, but still useful as a low-noise, motion artifact-free GT to compare reconstruction results against. The overall process of obtaining the data necessary for a GT is very time consuming. A sufficient number of OCT scans needs to be acquired, which have to be in focus, suffer from a minimal amount of motion artifacts, have similar alignment, and are scanned using orthogonal scan patterns. Only five to six scans could be acquired in a single sitting due to eye fatigue of the volunteer. Of 32 scans that were acquired over the course of a week, six were suitable for the generation of a GT.

\begin{figure}[htbp]
    \changed{M3.5}{\link{C3}}{}
    \centering
    \includegraphics[width=\columnwidth]{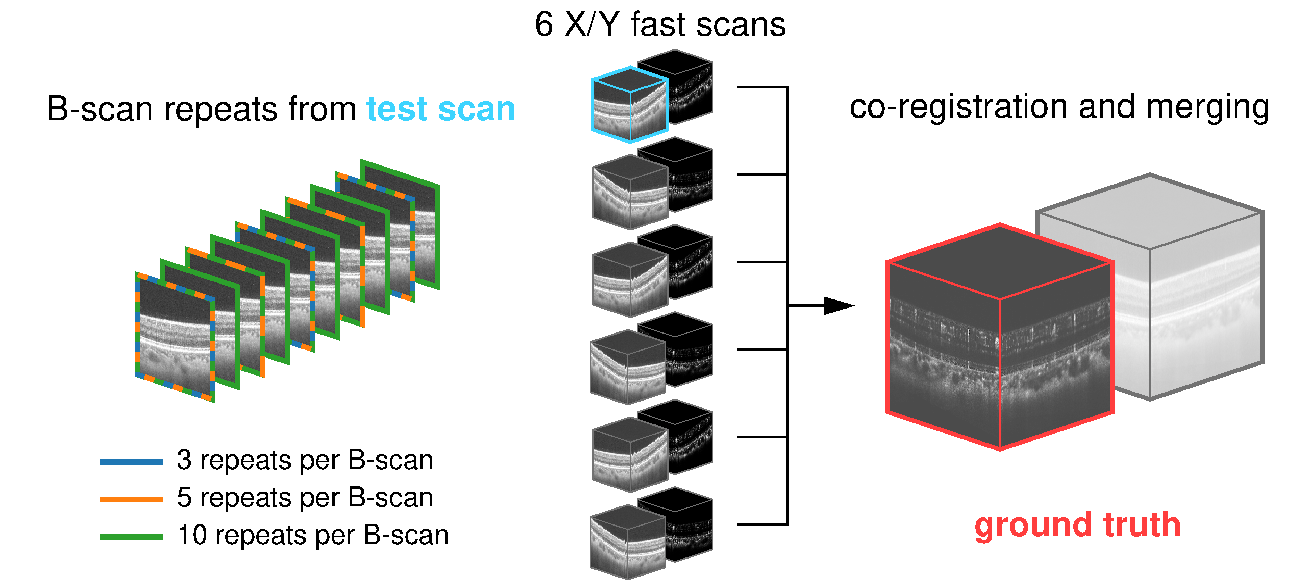}
    \caption{Use of test data during the evaluation. Six OCT scans were acquired for testing (center). OCTA volumes were computed from the test scans and structural/angiographic volume pairs were co-registered and merged. The merged OCTA volume serves as ground truth (red, right). Every one of the six test scans was acquired using ten repeats per B-scan. One of these scans is used as test scan (blue). Of the ten repeats, reconstruction was evaluated using three, five, and all ten repeats per B-scan (left).}
    \label{fig:data_flow}
\end{figure}
The data were acquired from a healthy 24 year old male volunteer. The OCT system used is a ultrahigh speed swept source OCT research prototype developed at the Massachusetts Institute of Technology and used by the New England Eye Center at Tufts Medical Center in Boston. The scanned field size is $3\times 3$ mm with 400 A-scans by 400 B-scans. The system runs at a scan rate of 600 kHz and scanned each B-scan ten times with an inter-scan time of \textasciitilde0.83 \textmu s. Power on cornea is \textasciitilde4.5 mW. Axial and lateral resolution are \textasciitilde8 \textmu m and \textasciitilde20 \textmu m respectively. Six scans were acquired with alternating scan patterns. Three scans were acquired using an X-fast scan pattern while the other three were acquired using a Y-fast scan pattern. Having both X and Y-fast scans ensures the necessary scan orthogonality for the motion correction algorithm. During each scan, every B-scan was repeatedly scanned ten times. This allows to compute OCTA data using the angiography formulas. \figurename~\ref{fig:data_flow} illustrates the data used and the relationship between the GT and test data. The six scans are depicted in the center with the scan used for testing highlighted blue. All six were co-registered and merged to obtain the OCTA GT visible on the right. Because the test scan, like the other five scans, was scanned with ten B-scan repeats, we decided to test reconstruction using  three, five and all ten B-scans, which is illustrated on the left. In the case of three repeats, the first, fifth, and ninth repeat were used. For five repeats, every second repeat was used.

To compare the reconstructed OCTA data with the GT, en face projections of the respective volumes were generated. OCTA data are commonly viewed as en face projections, because it allows clinicians to see the retinal vasculature or the choriocapillaris. In order to obtain clean OCTA en face projections, the volumes were segmented using segmentation software developed by Schottenhamml et al.\@ \cite{Schottenhamml2018}. The space between the outer nerve fiber layer and the inner plexiform layer was projected using a 98\textsuperscript{th} percentile projection. This corresponds to the superficial capillary plexus in the retina. In the case of AD, a background thresholding is performed to avoid excess noise in the en face image. The en face projections of the reconstructed OCTA volume can then be compared to the en face projection of the GT and peak signal-to-noise ratio (PSNR) and structural similarity index (SSIM) is computed. We chose SSIM in addition to PSNR because it is not based on the absolute error, but is based on the human visual perception of changes in structure \cite{Wang2004}.

\subsection{Patient Data}
In order to not only test on the scans of one eye, we decided to apply the reconstruction algorithm to existing scans of patient eyes. These data were not acquired with the generation of a GT in mind, which is very time-consuming and labor-intensive, thus the reconstruction results can only be evaluated qualitatively. But it allows to observe reconstruction results on more OCTA scans where there is also pathology present.

Patient 1 is male and was 41 years old at the time of imaging and exhibits severe non-proliferative diabetic retinopathy (NPDR) in connection with diabetic macular edema. Patient 2 is female and was 88 years old at the time of imaging and shows dry age-related macular degeneration and severe NPDR in both eyes. The system used was a 400 kHz Swept Source OCT device. Field size is $3 \times 3 \text{mm}$ with 500 A-scans by 500 B-scans. Every B-scan was repeatedly scanned five times with an inter-scan of \textasciitilde1.5 ms. Power on cornea was \textasciitilde1.8 mW \cite{Choi2013, Choi2015}.

\section{Results}
\begin{figure}[htbp]
    \centering
    \includegraphics[width=\textwidth]{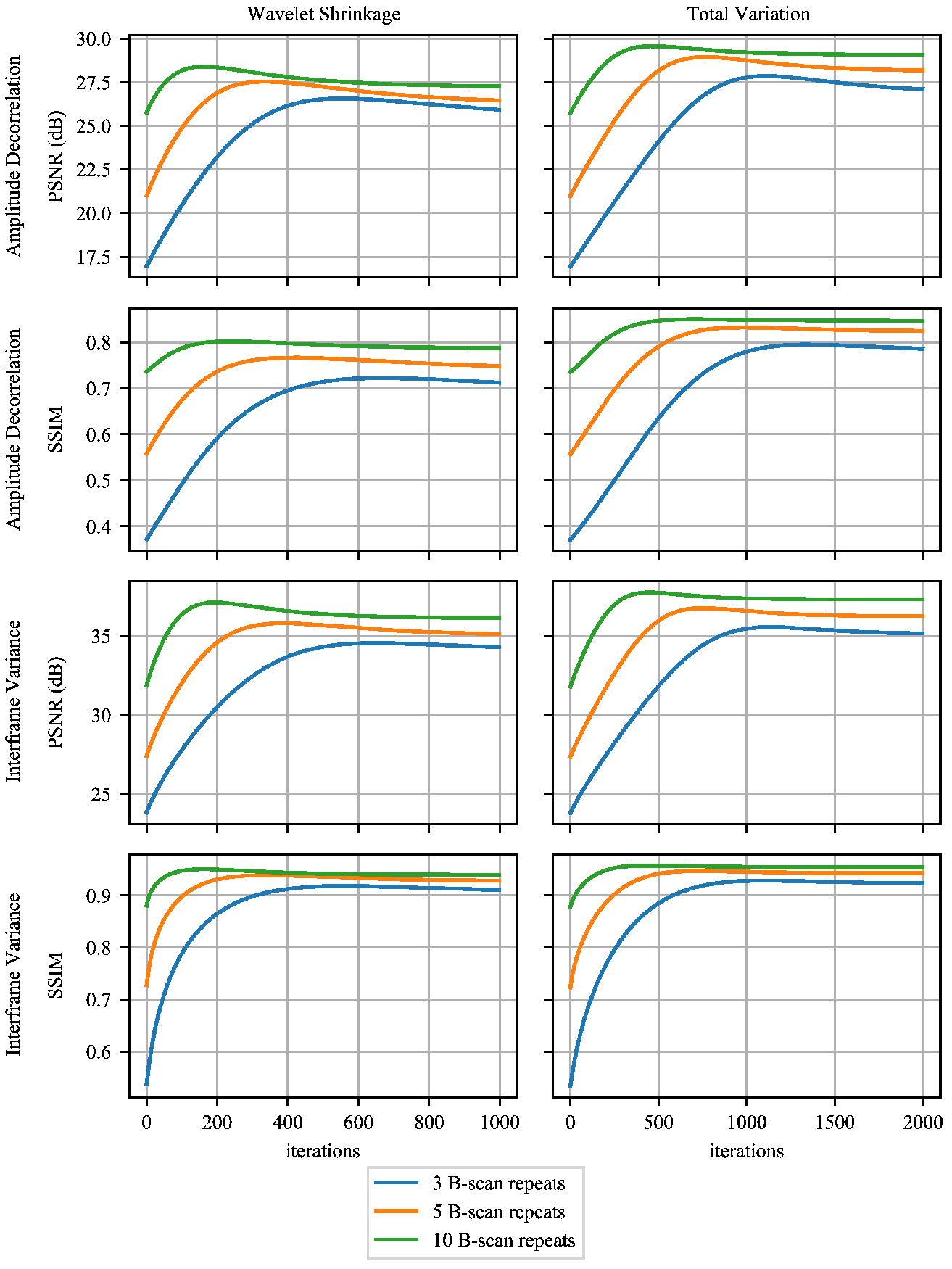}
    \caption{Peak signal-to-noise ratio (PSNR) and structural similarity (SSIM) during reconstruction. The color indicates whether three, five, or ten repeated B-scans were used for reconstruction. The rows show PSNR for amplitude decorrelation and interframe variance, the columns indicate the regularizer.}
    \label{fig:psnr}
\end{figure}
\begin{figure}[htbp]
    \centering
    \includegraphics[width=.955\columnwidth]{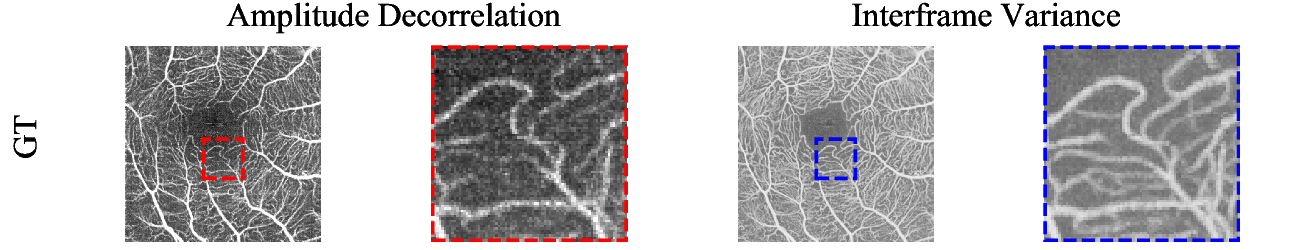}
    \includegraphics[width=.955\columnwidth]{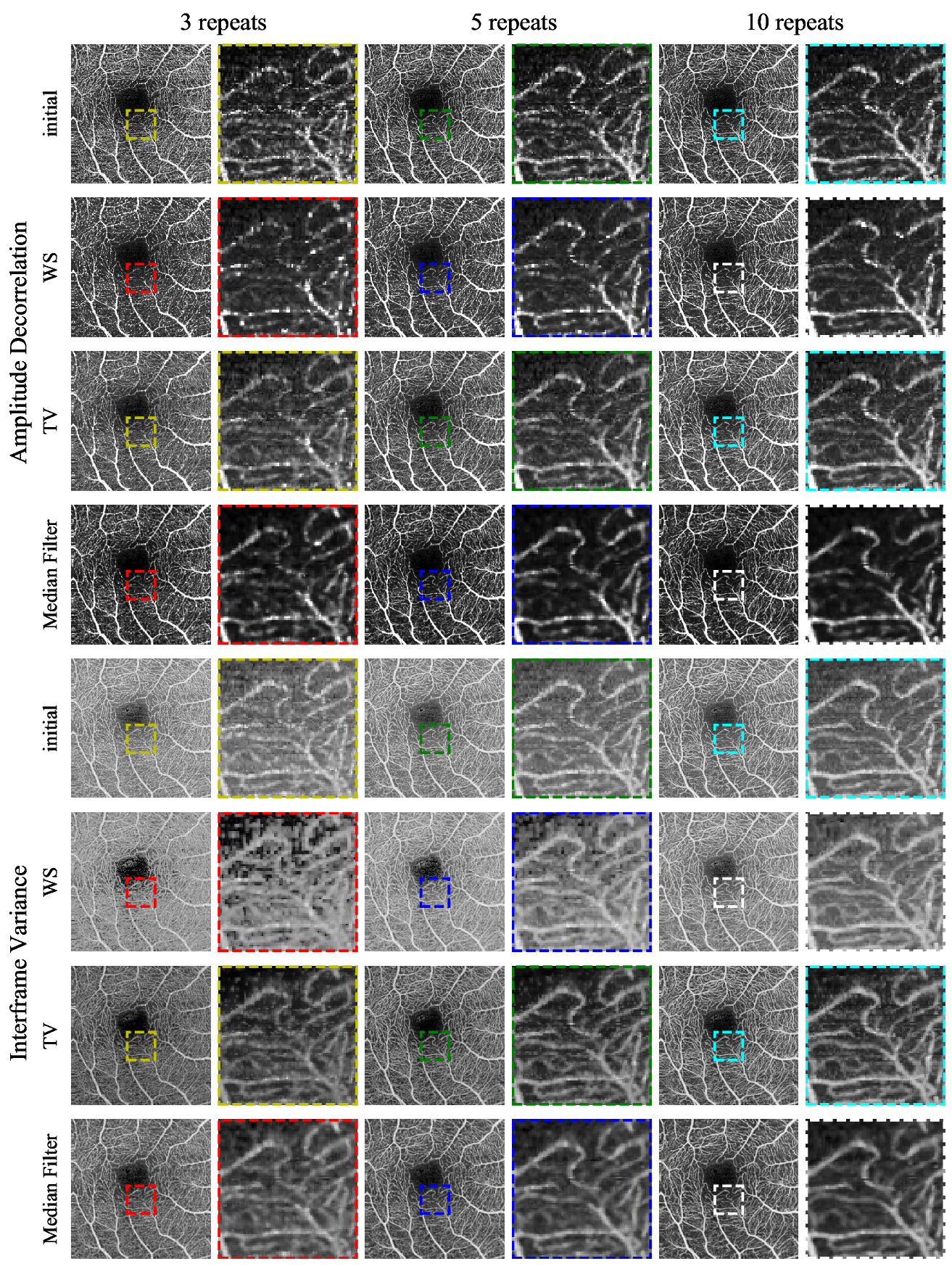}
    \caption{En face projections of the GT and reconstruction results. More detailed areas are shown in enlargements next to the en face images. Rows are grouped by AD and IF, and subdivided by initial OCTA, wavelet shrinkage and total variation reconstructions and median filter for comparison. Columns are grouped by the number of scan repeats used.}
    \label{fig:en_face_ad_ifv}
\end{figure}
\figurename~\ref{fig:psnr} shows PSNR and SSIM metrics for the reconstruction tests that compare against the GT. The colors indicate how many B-scan repeats were used during reconstruction. The two top rows show AD results and the bottom row IFV. The plots in the left column show results using \replaced{M3.6}{\link{C3}}{WS regularization}{wavelet shrinkage regularization (WS)} and the plots in the right column show \replaced{M3.7}{\link{C3}}{TV minimization}{total variation minimization (TV)} results. For AD and IFV, PSNR results are shown first, followed by SSIM results directly below. Reconstruction using ten scan repeats consistently performs best, although the largest improvements can be observed when using three repeats. IFV shows better results using both PSNR and SSIM. \figurename~\ref{fig:en_face_ad_ifv} shows 98\textsuperscript{th} percentile en face projections based on AD and IFV. The first row shows GT projections for AD and IFV. Enlarged sections of each en face image are shown to the right of the respective en face image. Rows are grouped by AD and IFV first, then by the initial OCTA volume, then \replaced{M3.8}{\link{C3}}{wavelet shrinkage}{WS} and \replaced{M3.9}{\link{C3}}{total variation}{TV} reconstruction results and, for comparison, median filtered results of the initial OCTA volumes using a 3D median filter using a $3\times3\times3$ kernel. Columns are grouped by the number of scan repeats used.
Furthermore, Tab.~\ref{tab:results} shows PSNR and SSIM results in more detail, including the 3D median filter results.
\begin{table}[t]
    \centering
    \begin{tabular}{cc cc cc cc cc}
           \hline
           & Repeats & \multicolumn{2}{c}{Initial OCTA} & \multicolumn{2}{c}{Wavelet Shrinkage} & \multicolumn{2}{c}{Total Variation} & \multicolumn{2}{c}{Median Filter}  \\
           && PSNR & SSIM & PSNR & SSIM & PSNR & SSIM & PSNR & SSIM \\
         \hline
            & 3 & 16.92 & 0.36 & 27.63 & 0.78 & \textbf{29.43} & \textbf{0.84} & 27.18 & 0.73\\
         AD & 5 & 20.96 & 0.55 & 28.64 & 0.81 & \textbf{30.29} & \textbf{0.86} & 27.25 & 0.73\\
            & 10 & 25.70 & 0.73 & 29.30 & 0.83 & \textbf{30.50} & \textbf{0.86} & 26.95 & 0.72\\
         \hline
             & 3 & 23.78 & 0.54 & 34.55 & 0.91 & 35.56 & \textbf{0.93} & \textbf{36.03} & \textbf{0.93}\\
         IFV & 5 & 27.33 & 0.72 & 35.81 & 0.93 & \textbf{36.76} & \textbf{0.95} & 36.55 & 0.93\\
             & 10 & 31.78 & 0.88 & 37.12 & 0.95 & \textbf{37.76} & \textbf{0.96} & 36.71 & 0.92\\

         \hline
    \end{tabular}
    \caption{Reconstruction results using the GT for comparison. The top rows show the results for AD and the bottom rows for IFV. Results for three, five, and ten scan repeats are sorted by row. PSNR and SSIM measures are provided for the initial OCTA volume, wavelet shrinkage and total variation reconstruction results and the median filtered initial volume. The final reported value corresponds to the final iteration result of the reconstruction. The best PSNR and SSIM values in each row are marked in bold.}
    \label{tab:results}
\end{table}
\begin{figure}[htbp]
    \centering\includegraphics[width=\columnwidth]{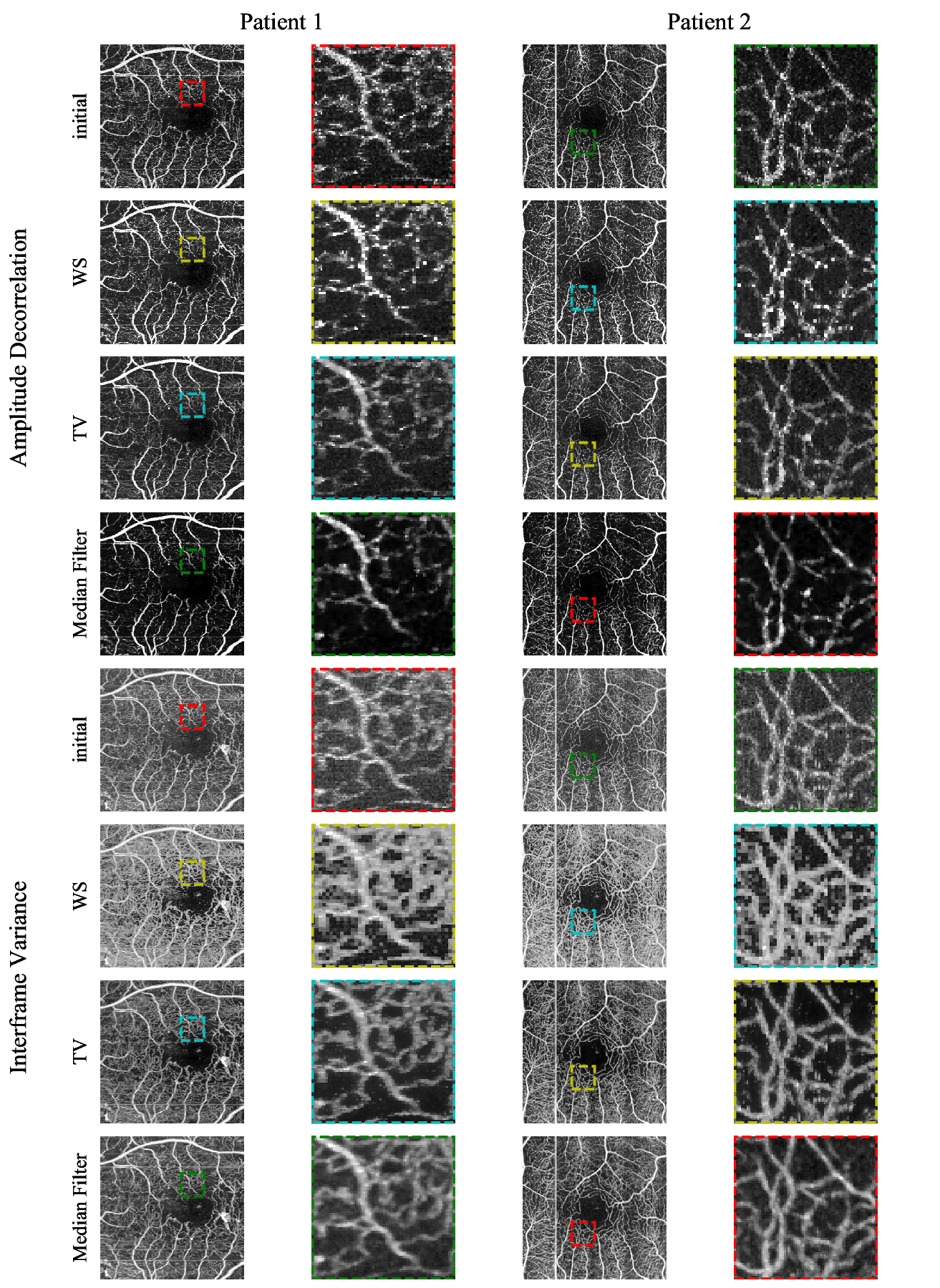}
    \caption{En face projections of reconstructed patient data. Rows are grouped by AD and IFV, then by the initial OCTA volume, followed by wavelet shrinkage and total variation reconstructions. Columns are grouped by patient, enlarged areas are directly to the right of their projections.}
    \label{fig:en_face_patients}
\end{figure}
Results for SV are included in the supplementary document.


\section{Discussion}
\figurename~\ref{fig:psnr} shows PSNR and SSIM over reconstruction iterations while Tab.~\ref{tab:results} shows the highest PSNR and SSIM values during reconstruction. A slight dipping of the curves during reconstruction might indicate over-regularization, but, as can be seen in \figurename~\ref{fig:psnr}, a decrease in PSNR does not necessarily correspond with a decrease in SSIM.

It is apparent that using more B-scan repeats, or samples, generally leads to better PSNR and SSIM while the initial OCTA and early reconstruction show larger discrepancies than the final results. The raw, unreconstructed en face projections for AD have PSNRs of \textasciitilde16.92 dB, \textasciitilde20.96 dB, and \textasciitilde25.70 dB for three, five, and ten repeats respectively. Reconstruction results using WS regularization show PSNRs of \textasciitilde27.63 dB, \textasciitilde28.64 dB, and \textasciitilde29.30 dB, while TV regularized PSNRs are \textasciitilde29.43 dB, \textasciitilde30.29 dB, and \textasciitilde30.50 dB. To put it differently, using more repeats generally leads to better results but with diminishing returns during reconstruction. This is an important consideration, since more B-scan repeats give more information to compute OCTA from, while longer scan times make the entire scan more susceptible to motion artifacts. \deleted{M2.10}{\link{C2}}{This result is expected and in line with CS.} This is visually supported by the en face images in figure \figurename~\ref{fig:en_face_ad_ifv} where there is hardly a discernible difference between the projections for 5 and 10 repeats, while the difference to 3 repeats is easily apparent in the form of more noise. Commercial systems use two or three B-scan repeats where the benefits of reconstruction is highest, but testing reconstruction using more repeats shows that the algorithm is working correctly. One observation of interest is that SSIM of the median filtered result slightly decreases from five to ten repeats. 

IFV results achieve both higher PSNR in range of \textasciitilde34 dB to \textasciitilde37.7 dB and SSIM of more than 0.9 when compared to AD with \textasciitilde27.6 dB to \textasciitilde30.50 dB and \textasciitilde0.8 respectively. Direct comparisons between IFV and AD results are of limited use however. Because of the normalization which is part of the AD computation, its results tend to be very noisy. Background thresholding can compensate for this, but can erase detail such as small capillaries. Comparing AD and IFV GT in \figurename~\ref{fig:en_face_ad_ifv} shows, despite background thresholding for AD, that more noise is visible in intercapillary areas and less capillary details. For this reason, IFV en face images start out at better PSNR and SSIM to begin with than AD images.

Comparing WS to TV regularization is more meaningful though. The measurements in Tab.~\ref{tab:results} consistently show better results when using TV regularization. In the case of AD, PSNR and SSIM results range from \textasciitilde27.63 dB to \textasciitilde30.50 dB and \textasciitilde0.78 to \textasciitilde0.86. For IFV, these results are much closer with \textasciitilde37.12 dB to \textasciitilde37.76 and \textasciitilde0.95 to \textasciitilde0.96. When looking at the images in \figurename~\replaced{M3.10}{\link{C3}}{\ref{eq:ifv}}{\ref{fig:en_face_ad_ifv}} the qualitative differences between WS and TV regularization can be easily seen. The median filtered PSNR and SSIM results, however, do not show an advantage when using IFV. Median filter PSNR is higher than 36 dB for all numbers of repeats. In the case of three repeats, both PSNR and SSIM indicate that the median filter performs better than TV or WS regularized reconstruction. When computing AD data though, both WS and TV regularization have better PSNR and SSIM values than the median filter comparison.

\figurename~\ref{fig:en_face_ad_ifv} shows en face projections of the GT and reconstruction results. GT projections for both AD and IFV show that those are not completely noise-free. Co-registering and merging six volumes does not result in a perfect gold standard GT. When comparing to the reconstruction results below, it becomes evident that the GT displays significantly less noise and can still serve as a silver-standard GT. First, we take a look at the AD en face projections in \figurename~\ref{fig:en_face_ad_ifv}. The initial AD projections show significant amounts of noise which continually decrease the more repeats are used despite background thresholding. WS regularization seems to lead to less noisy intercapillary areas, although blood vessels are more faint, when compared to TV regularization. WS regularization also leads to isolated bright pixels in blood vessels. TV regularization my preserver more noise in the intercapillary areas compared to WS, but fine capillaries remain more visible. This is consistent with the results in Tab.~\ref{tab:results}. While, in the case of ten repeats, it can be argued that WS regularization provides improvements over the initial projection, the same is less obvious for the TV regularization. Although the median filtered results seem to provide more contrast between vessels and intercapillary areas at first glance, vessels are blurred and fine capillaries are harder to discern and less continuous.

\figurename~\ref{fig:en_face_patients} shows the en face results for the two patient scans. Again, WS regularization used in conjunction with AD leads to brightened sections in some blood vessels. TV regularization makes the best impression, while the median-filtered comparison is exceedingly blurry with the loss of finer vessels. With IFV data, WS regularization increases contrast but leads to irregular, jagged borders along the vessels. TV regularization looks best, with enhanced contrast and a better delineation between vessels and the areas in-between, while the median-filtered comparison is more blurry.

We do not provide a formal proof for the convergence of our algorithm, but the plots in \figurename~\ref{fig:psnr} indicate that the algorithm converges after a certain number of iterations depending on the regularizer and parameters. When used in practice, there are no GT data available during reconstruction of OCTA volumes. One way of determining whether convergence has been reached, is to compare the current reconstruction result with the initial OCTA data and stop reconstruction once an error metric such as mean squared error between the two stopped changing.

\section{Conclusion and Outlook}
In this paper, we present a novel MAP \replaced{M1.8}{\link{C1}}{estimate based data consistency term for the reconstruction of OCTA volumes derived from Ploner et al.\@ previous maximum likelihood estimates for OCTA. This is, to the best of our knowledge, the first application of Bayesian statistics for the reconstruction of OCTA images.}{estimation based data consistency term for the reconstruction of OCTA volumes. It extends the MLE of Ploner at al.\@ with regularization terms, thereby harnessing the inverse problem formulation's potential.} We present OCTA reconstruction results for AD and IFV data and compare them to GT OCTA data merged from six co-registered OCT scans.
Generally, TV regularization outperforms WS both quantitatively and qualitatively. Contrast between vessels and areas in-between seems significantly increased, with only very minor blurring. While median-filtering also leads to increases in contrast, those en face images are significantly more blurry. TV reconstruction achieves PSNR of more \textasciitilde29.43 dB for AD and more than ~\textasciitilde35.56 dB for IFV. Except in the case of three scan repeats using IFV, TV regularized reconstruction outperforms 3D median filtering. Qualitative results show a clear reduction in noise in the FAZ and intercapillary areas when compared to the state-of-the-art. Although this is a simple iterative algorithm, it shows significant improvement for OCTA image data and \deleted{M1.9}{\link{C1}}{opens up new ways for the reconstruction of OCTA images. This can be achieved by allowing the use of a large number of CS algorithms already established in other fields of medical imaging. The proposed reconstruction algorithm} opens OCTA up to more sophisticated optimization techniques, such as alternating direction method of multipliers (ADMM). \changed{M1.10}{\link{C1}}{The method presented in this paper is, to the best of our knowledge, the first application of Bayesian statistics for the reconstruction of OCTA images.}

\section*{Funding}
Deutsche Forschungsgemeinschaft (DFG) (MA 4898/12-1); National Institutes of Health (NIH) (5-R01-EY011289-31); Champalimaud Vision Award; Beckman-Argyros Award in Vision Research; Macula Vision Research Foundation (MVRF); Massachusetts Lions Clubs


\section*{Disclosures}
NKW: Optovue (C), Carl Zeiss Meditec (F), Heidelberg (F), Nidek (F). JGF: Optovue (I, P), Topcon (F)\\

\noindent
See Supplement 1 for supporting content.

\bibliography{ms}

\begin{thebibliography}{10}
\newcommand{\enquote}[1]{``#1''}

\bibitem{Choi2015}
W.~Choi, E.~M. Moult, N.~K. Waheed, M.~Adhi, B.~Lee, C.~D. Lu, T.~E. de~Carlo,
  V.~Jayaraman, P.~J. Rosenfeld, J.~S. Duker, and J.~G. Fujimoto,
  \enquote{{Ultrahigh-Speed, Swept-Source Optical Coherence Tomography
  Angiography in Nonexudative Age-Related Macular Degeneration with Geographic
  Atrophy},} {\protect\JournalTitle{Ophthalmology}} \textbf{122}, 2532--2544
  (2015).

\bibitem{Novais2016}
E.~A. Novais, M.~Adhi, E.~M. Moult, R.~N. Louzada, E.~D. Cole, L.~Husvogt,
  B.~Lee, S.~Dang, C.~V. Regatieri, A.~J. Witkin, C.~R. Baumal, J.~Hornegger,
  V.~Jayaraman, J.~G. Fujimoto, J.~S. Duker, and N.~K. Waheed,
  \enquote{{Choroidal Neovascularization Analyzed on Ultrahigh-Speed
  Swept-Source Optical Coherence Tomography Angiography Compared to
  Spectral-Domain Optical Coherence Tomography Angiography},}
  {\protect\JournalTitle{American Journal of Ophthalmology}} \textbf{164},
  80--88 (2016).

\bibitem{Jia2012}
Y.~Jia, O.~Tan, J.~Tokayer, B.~Potsaid, Y.~Wang, J.~J. Liu, M.~F. Kraus,
  H.~Subhash, J.~G. Fujimoto, J.~Hornegger, and D.~Huang,
  \enquote{{Split-spectrum amplitude-decorrelation angiography with optical
  coherence tomography},} {\protect\JournalTitle{Optics Express}} \textbf{20},
  4710 (2012).

\bibitem{Mariampillai2008}
A.~Mariampillai, B.~A. Standish, E.~H. Moriyama, M.~Khurana, N.~R. Munce,
  M.~K.~K. Leung, J.~Jiang, A.~E. Cable, B.~C. Wilson, I.~A. Vitkin, and
  V.~X.~D. Yang, \enquote{{Speckle variance detection of microvasculature using
  swept-source optical coherence tomography},} {\protect\JournalTitle{Optics
  Letters}} \textbf{33}, 1530--1532 (2008).

\bibitem{Lustig2007}
M.~Lustig, D.~Donoho, and J.~M. Pauly, \enquote{{Sparse MRI: The application of
  compressed sensing for rapid MR imaging},} {\protect\JournalTitle{Magnetic
  Resonance in Medicine}}  (2007).

\bibitem{Sidky2008}
E.~Y. Sidky and X.~Pan, \enquote{{Image reconstruction in circular cone-beam
  computed tomography by constrained, total-variation minimization},}
  {\protect\JournalTitle{Physics in Medicine and Biology}} \textbf{53},
  4777--4807 (2008).

\bibitem{Wu2012}
H.~Wu, A.~Maier, R.~Fahrig, and J.~Hornegger, \enquote{{Spatial-temporal total
  variation regularization (STTVR) for 4D-CT reconstruction},} in \emph{Medical
  Imaging 2012: Physics of Medical Imaging,}  vol. 8313 N.~J. Pelc, R.~M.
  Nishikawa, and B.~R. Whiting, eds. (International Society for Optics and
  Photonics, 2012).

\bibitem{Amrehn2015}
M.~Amrehn, A.~Maier, F.~Dennerlein, and J.~Hornegger, \enquote{{Portability of
  TV-Regularized Reconstruction Parameters to Varying Data Sets},} in
  \emph{Bildverarbeitung f{\"{u}}r die Medizin 2015,}  H.~Handels, T.~M.
  Deserno, H.-P. Meinzer, and T.~Tolxdorff, eds. (Springer Vieweg, Berlin,
  Heidelberg, Berlin, Heidelberg, 2015), pp. 131--136.

\bibitem{Stromer2016}
D.~Stromer, M.~Amrehn, Y.~Huang, P.~Kugler, S.~Bauer, G.~Lauritsch, and
  A.~Maier, \enquote{{Comparison of SART and ETV reconstruction for increased
  C-arm CT volume coverage by proper detector rotation in liver imaging},} in
  \emph{2016 IEEE 13th International Symposium on Biomedical Imaging (ISBI),}
  (IEEE, 2016), pp. 589--592.

\bibitem{Huang2016}
Y.~Huang, G.~Lauritsch, M.~Amrehn, O.~Taubmann, V.~Haase, D.~Stromer, X.~Huang,
  and A.~Maier, \enquote{{Image Quality Analysis of Limited Angle Tomography
  Using the Shift-Variant Data Loss Model},} in \emph{Bildverarbeitung
  f{\"{u}}r die Medizin 2016,}  T.~Tolxdorff, T.~M. Deserno, H.~Handels, and
  H.-P. Meinzer, eds. (Springer Vieweg, Berlin, Heidelberg, Berlin, Heidelberg,
  2016), pp. 277--282.

\bibitem{Wong2010}
A.~Wong, A.~Mishra, K.~Bizheva, and D.~A. Clausi, \enquote{{General Bayesian
  estimation for speckle noise reduction in optical coherence tomography
  retinal imagery},} {\protect\JournalTitle{Optics Express}} \textbf{18}, 8338
  (2010).

\bibitem{Ploner2018}
S.~B. Ploner, C.~Riess, J.~Schottenhamml, E.~M. Moult, N.~K. Waheed, J.~G.
  Fujimoto, and A.~Maier, \enquote{{A Joint Probabilistic Model for Speckle
  Variance, Amplitude Decorrelation and Interframe Variance (IFV) Optical
  Coherence Tomography Angiography},} in \emph{Bildverarbeitung f{\"{u}}r die
  Medizin 2018,}  A.~Maier, T.~M. Deserno, H.~Handels, K.~H. Maier-Hein,
  C.~Palm, and T.~Tolxdorff, eds. (Springer Berlin Heidelberg, Berlin,
  Heidelberg, 2018), 211279, pp. 98--102.

\bibitem{Zeng2010}
G.~L. Zeng, \emph{{Medical Image Reconstruction}} (Springer Berlin Heidelberg,
  Berlin, Heidelberg, 2010), 1st ed.

\bibitem{Manhart2013}
M.~T. Manhart, M.~Kowarschik, A.~Fieselmann, Y.~Deuerling-Zheng, K.~Royalty,
  A.~Maier, and J.~Hornegger, \enquote{{Dynamic Iterative Reconstruction for
  Interventional 4-D C-Arm CT Perfusion Imaging},} {\protect\JournalTitle{IEEE
  Transactions on Medical Imaging}} \textbf{32}, 1336--1348 (2013).

\bibitem{Neukirchen2010}
C.~Neukirchen, M.~Giordano, and S.~Wiesner, \enquote{{An iterative method for
  tomographic x-ray perfusion estimation in a decomposition model-based
  approach},} {\protect\JournalTitle{Medical Physics}} \textbf{37}, 6125--6141
  (2010).

\bibitem{Chang2000}
S.~G. Chang, B.~Yu, and M.~Vetterli, \enquote{{Adaptive wavelet thresholding
  for image denoising and compression},} {\protect\JournalTitle{IEEE
  Transactions on Image Processing}} \textbf{9}, 1532--1546 (2000).

\bibitem{Chambolle2004}
A.~Chambolle, \enquote{{An Algorithm for Total Variation Minimization and
  Applications},} {\protect\JournalTitle{Journal of Mathematical Imaging and
  Vision}} \textbf{20}, 89--97 (2004).

\bibitem{Landweber1951}
L.~Landweber, \enquote{{An Iteration Formula for Fredholm Integral Equations of
  the First Kind},} {\protect\JournalTitle{American Journal of Mathematics}}
  \textbf{73}, 615--624 (1951).

\bibitem{VanDerWalt2011}
S.~{Van Der Walt}, S.~C. Colbert, and G.~Varoquaux, \enquote{{The NumPy array:
  A structure for efficient numerical computation},}
  {\protect\JournalTitle{Computing in Science and Engineering}} \textbf{13},
  22--30 (2011).

\bibitem{VanDerWalt2014}
S.~{Van Der Walt}, J.~L. Sch{\"{o}}nberger, J.~Nunez-Iglesias, F.~Boulogne,
  J.~D. Warner, N.~Yager, E.~Gouillart, and T.~Yu, \enquote{{Scikit-image:
  Image processing in python},} {\protect\JournalTitle{PeerJ}} \textbf{2014},
  e453 (2014).

\bibitem{Mayer2012}
M.~a. Mayer, A.~Borsdorf, M.~Wagner, J.~Hornegger, C.~Y. Mardin, and R.~P.
  Tornow, \enquote{{Wavelet denoising of multiframe optical coherence
  tomography data},} {\protect\JournalTitle{Biomedical Optics Express}}
  \textbf{3}, 572 (2012).

\bibitem{Kraus2012}
M.~F. Kraus, B.~Potsaid, M.~a. Mayer, R.~Bock, B.~Baumann, J.~J. Liu,
  J.~Hornegger, and J.~G. Fujimoto, \enquote{{Motion correction in optical
  coherence tomography volumes on a per A-scan basis using orthogonal scan
  patterns},} {\protect\JournalTitle{Biomedical Optics Express}} \textbf{3},
  1182 (2012).

\bibitem{Kraus2014}
M.~F. Kraus, J.~J. Liu, J.~Schottenhamml, C.-L. Chen, A.~Budai, L.~Branchini,
  T.~Ko, H.~Ishikawa, G.~Wollstein, J.~S. Schuman, J.~S. Duker, J.~G. Fujimoto,
  and J.~Hornegger, \enquote{{Quantitative 3D-OCT motion correction with tilt
  and illumination correction, robust similarity measure and regularization},}
  {\protect\JournalTitle{Biomedical Optics Express}} \textbf{5}, 2591 (2014).

\bibitem{ploner20183}
S.~B. Ploner, M.~F. Kraus, L.~Husvogt, E.~Moult, A.~Y. Alibhai,
  J.~Schottenhamml, T.~Geimer, C.~Rebhun, B.~Lee, C.~R. Baumal, N.~K. Waheed,
  J.~S. Duker, J.~G. Fujimoto, and A.~Maier, \enquote{{3-D OCT Motion
  Correction Efficiently Enhanced with OCT Angiography},} in
  \emph{Investigative Ophthalmology \& Visual Science,}  vol.~59 (The
  Association for Research in Vision and Ophthalmology, 2018), p. 3922.

\bibitem{Ploner2020}
S.~B. Ploner, M.~F. Kraus, E.~M. Moult, L.~Husvogt, J.~Schottenhamml, A.~Y.
  Alibhai, N.~K. Waheed, J.~S. Duker, J.~G. Fujimoto, and A.~K. Maier,
  \enquote{{Efficient and high accuracy 3-D OCT angiography motion correction
  in pathology},}  (2020). {arXiv:2010.06931}.

\bibitem{Schottenhamml2018}
J.~Schottenhamml, E.~M. Moult, E.~A. Novais, M.~F. Kraus, B.~K. Lee, W.~Choi,
  S.~B. Ploner, L.~Husvogt, C.~D. Lu, P.~Yiu, P.~J. Rosenfeld, J.~S. Duker,
  A.~K. Maier, N.~Waheed, and J.~G. Fujimoto, \enquote{{OCT-OCTA
  Segmentation},} in \emph{Bildverarbeitung f{\"{u}}r die Medizin 2018,}
  A.~Maier, T.~M. Deserno, H.~Handels, K.~Maier-Hein, P.~C., and T.~Tolxdorff,
  eds. (Springer Vieweg, Berlin, Heidelberg, Berlin, Heidelberg, 2018), p. 284.

\bibitem{Wang2004}
Z.~Wang, A.~C. Bovik, H.~R. Sheikh, and E.~P. Simoncelli, \enquote{{Image
  quality assessment: From error visibility to structural similarity},}
  {\protect\JournalTitle{IEEE Transactions on Image Processing}} \textbf{13},
  600--612 (2004).

\bibitem{Choi2013}
W.~Choi, B.~Potsaid, V.~Jayaraman, B.~Baumann, I.~Grulkowski, J.~J. Liu, C.~D.
  Lu, A.~E. Cable, D.~Huang, J.~S. Duker, and J.~G. Fujimoto,
  \enquote{{Phase-sensitive swept-source optical coherence tomography imaging
  of the human retina with a vertical cavity surface-emitting laser light
  source},} {\protect\JournalTitle{Optics Letters}} \textbf{38}, 338--340
  (2013).

\end{thebibliography}

\end{document}